\documentclass[aps,prd,superscriptaddress,eqsecnum,nofootinbib,preprintnumbers,showpacs,showkeys]{revtex4}
\usepackage{graphicx}
\usepackage{euscript,amssymb}
\usepackage{amsfonts}
\usepackage{amsmath}
\usepackage{amssymb}
\usepackage{graphicx}
\usepackage{euscript}

\begin{document}

\title{\hspace*{0.5cm} Phantom-like behaviour in dilatonic brane-world scenario with induced gravity}

\author{Mariam Bouhmadi-L\'{o}pez}
\email{mariam.bouhmadi@fisica.ist.utl.pt}
\affiliation{Centro Multidisciplinar de Astrof\'{\i}sica - CENTRA, Departamento de F\'{\i}sica, Instituto Superior T\'ecnico, Av. Rovisco Pais 1,
1049-001 Lisboa, Portugal}


\begin{abstract}

The  Dvali, Gabadadze and Porrati (DGP) model has a self-accelerating solution, the positive  branch, where the brane is asymptotically de Sitter. 
A de Sitter space-time can be seen as a boundary between quintessence-like behaviour and phantom-like behaviour. 
We show that in a 5D dilatonic bulk, where the dilaton has an exponential potential,  with an induced gravity term on the brane, whose matter content corresponds 
only to vacuum energy,  the positive branch solution  undergoes  
a phantom-like stage where it faces a curvature singularity in its infinite future. The singularity can be interpreted as the  ``big rip'' singularity pushed towards an infinite future cosmic time. The phantom-like behaviour on the brane occurs without violating the null energy condition.
There is another solution, the negative branch, where the brane  
can undergo an early-epoch (transient) inflationary phase  induced by the dilaton field.

\end{abstract}

\date{\today}

\maketitle

\section{introduction}

The supernova Ia (SNIa) observations \cite{SNIa} and the cosmic microwave
background (CMB) anisotropy data \cite{Spergel:2003cb} suggest  that the
expansion of our universe seems to be accelerating.
A possible explanation for this evolution is the usual
vacuum energy represented by a cosmological constant
providing a negative pressure \cite{3b}. However, the observational value of $\Lambda$ is about
$120$ orders of magnitude smaller than that established from field
theory methods \cite{3b}.  So far, alternative phenomenological models have been proposed to describe the late-time acceleration of the universe \cite{Copeland:2006wr}.

One approach is to consider an  effective dark energy component  in the energy momentum tensor. For example, this component can be described by a scalar field  as  in quintessence models \cite{Wetterich:fm} or tachyonic models \cite{Gibbons:2002md}. Dark energy  can also be described effectively by a perfect fluid with a linear equation of state or a more general barotropic equation of state like in Chaplygin gas models \cite{Kamenshchik}.  Motivated initially by SNIa observations, phantom energy  models \cite{Caldwell:2003vq,Caldwell:1999ew,Bouhmadi-Lopez:2004me,singularity} have also been proposed to account for the late-time acceleration of the universe. A recent analysis \cite{Rapetti:2004aa} of the current equation of state for dark energy  based on CMB anisotropies, SNIa and X-ray galaxy cluster data concluded that the current equation of state is compatible with a phantom-like behaviour of dark energy; i.e. $w$, the ratio of  the pressure and the energy density of dark energy, could be less than $-1$. A similar conclusion is reached in \cite{Percival:2007yw}; i.e. observational data do not seem incompatible with phantom-like behaviour of dark energy. 
In phantom energy models the null energy condition is not satisfied. Hence, the energy density is an increasing function of the scale factor in an expanding Friedmann-Lema\^{\i}tre-Robertson-Walker (FLRW) universe. This may lead to the occurrence of a big rip singularity in the future evolution of a phantom energy  dominated universe \cite{Caldwell:2003vq,Bouhmadi-Lopez:2004me,singularity}.

An alternative approach to account for the late-time acceleration of the universe is to consider a generalised Einstein theory of gravity like brane-world models (for  reviews, see Ref.~\cite{review}) where
the observable four-dimensional (4D) universe is a brane
(hyper-surface) embedded in a higher-dimensional space (bulk). For example, the Dvali, Gabadadze and Porrati (DGP) model \cite{Dvali:2000hr}  has a self-accelerating solution at late-time which is asymptotically de Sitter \cite{Deffayet,brane}. In this model the bulk is a 5D Minkowski space-time and the brane action contains an induced gravity term 
\cite{Dvali:2000hr,Deffayet,brane,other1,lambdadgp,ktt,Papantonopoulos:2004bm,Maeda:2003ar,Bouhmadi-Lopez:2004ys,Bouhmadi-Lopez:2004ax}
which is proportional to the 4D Ricci scalar curvature of the brane. Dilatonic brane-world models \cite{Chamblin:1999ya,Maeda:2000wr,dilatonicbw,Feinstein:2001xs} can also account for late-time acceleration of the brane through a quintessence-like behaviour  driven by a bulk dilaton \cite{Kunze:2001ji}. One aim of this paper is to show that a dilatonic brane-world model with an induced gravity term in the brane can mimic a phantom-like behaviour without including matter on the brane that violates the null  energy condition.

In an induced gravity brane-world model the Friedmann equation  has two solutions (depending on the embedding of the brane in the bulk \cite{Deffayet}). One of these solutions (the self-accelerating solution or the positive branch) can account for the late-time evolution of the universe \cite{Deffayet,brane}.  The other solution (the negative branch) can describe the early-time evolution of the universe\footnote{For an alternative use of the negative branch of  the DGP scenario see Ref.~\cite{lambdadgp}.} \cite{ktt,Papantonopoulos:2004bm,Maeda:2003ar,Bouhmadi-Lopez:2004ys,Bouhmadi-Lopez:2004ax} and in particular corresponds to a correction to Randall-Sundrum (RS) model \cite{RS}. In 
the dilatonic brane-world model with induced gravity that we will analyse, the negative branch may undergo a transient inflationary epoch.   

The layout of this paper is as follows. In section 2, we present our dilatonic brane-world model with induced gravity. The bulk scalar field potential is a Liouville potential. The matter content of the brane is coupled to the dilaton field. We deduce the modified Friedmann equation for both branches, the junction condition of the dilaton across the brane, which constrains the brane tension, and the energy balance on the brane. In section 3, we analytically derive the  solutions of a vacuum brane for both branches; i.e. whose matter content is described through the brane tension. We show that the brane tension has a phantom-like energy density behaviour on the positive branch in the sense that the brane tension grows as the brane expands. The brane hits a singularity in its future evolution which may be interpreted as a ``big rip'' singularity pushed towards an infinite cosmic time. The negative branch can undergo an early-epoch (transient) inflationary phase  induced by the dilaton field through the brane tension. Finally, we conclude and summarise in section 4.

\section{The model}

We consider a  brane, described by a 4D hyper-surface ($h$, metric g), embedded in a 5D bulk space-time ($\mathcal{B}$, metric 
$g^{(5)}$), whose action is given by 
\begin{eqnarray}
\mathcal{S} =  \frac{1}{\kappa_5^2}\int_{\mathcal{B}} d^5X\, \sqrt{-g^{(5)}}\;
\left\{\frac{1}{2}R[g^{(5)}]\;+\;\mathcal{L}_5\right\}
+ \int_{h} d^4X\, \sqrt{-g}\; \left\{\frac{1}{\kappa_5^2} K\;+\;\mathcal{L}_4 \right\}\,, \label{action}
\end{eqnarray}
where $\kappa_5^2$ is the 5D gravitational constant,
$R[g^{(5)}]$ is the scalar curvature in the bulk and $K$ the extrinsic curvature of the brane in the higher dimensional
bulk, corresponding to the York-Gibbons-Hawking boundary term \cite{YGH}.  

We consider that there is a dilaton  field $\phi$ living on the bulk and we choose $\phi$ to be dimensionless. 
Then, the 5D Lagrangian $\mathcal{L}_5$ can be written as
\begin{eqnarray}
\mathcal{L}_5=-\frac12 (\nabla\phi)^2 -V(\phi).
\end{eqnarray}

The 4D Lagrangian $\mathcal{L}_4$ corresponds to 
\begin{equation}
\mathcal{L}_4= \alpha {R}[g] -\lambda(\phi)+\Omega^{4}\mathcal{L}_m(\Omega^2 g_{\mu\nu}).
\label{L4}\end{equation}
The first term on the right hand side (rhs) of the previous equation corresponds to an induced gravity term \cite{Dvali:2000hr,Deffayet,brane,other1,lambdadgp,ktt,Papantonopoulos:2004bm,Maeda:2003ar,Bouhmadi-Lopez:2004ys,Bouhmadi-Lopez:2004ax},
where $R[g]$ is the scalar curvature of the induced metric on the
brane and $\alpha$ is  a positive parameter which  measures the strength of the induced gravity term 
and has dimensions of mass squared. The term $\mathcal{L}_m$ in Eq.~(\ref{L4}) describes  the matter content of the brane and $\lambda(\phi)$ is the brane tension.
We allow the brane matter content to  be  non-minimally coupled on the (5D) Einstein frame  but to be minimally coupled respect to a conformal metric 
${\tilde g}^{(5)}_{AB}=\Omega^2\;g^{(5)}_{AB}$, where $\Omega=\Omega(\phi)$.

We are mainly interested in the cosmology of a homogeneous and isotropic brane, with an induced gravity term \cite{Maeda:2003ar,Bouhmadi-Lopez:2004ys}, and embedded in a bulk with a dilaton field \cite{Chamblin:1999ya}. It is known that
for an expanding FLRW brane the unique bulk space-time in Einstein gravity (in vacuum) is a 5D Schwarzschild-anti de Sitter space-time 
\cite{BCG,MSM}. This property in principle cannot be extended  to a 5D dilatonic bulk.
On the other hand, we stress that the presence of an induced gravity term in the  brane-world scenario
affects the dynamics of the brane, through the junction conditions at the brane, and does not affect the bulk field equations.
Consequently, in order to study the  effect of an induced gravity term in a brane-world dilatonic model, it is possible to consider a bulk
corresponding to a dilatonic 5D space-time and later on impose the junction conditions at the brane. The junction 
conditions will then determine the dynamics of the brane and constrain the brane tension. This is the approach we will follow. 

\subsection{The bulk}

From now on, we will restrict our attention to a 5D dilatonic solution obtained by Feinstein et al \cite{Feinstein:2001xs} and analysed 
by Kunze and V\'azquez-Mozo in the context of quintessence brane-world models \cite{Kunze:2001ji}
\textit{without an induced gravity term on the brane}. The 5D dilatonic solution reads \cite{Kunze:2001ji}
\begin{equation}
ds^2_5=\frac{1}{\xi^2}r^{2/3(k^2-3)}dr^2 +r^2(-d{t}^2+\gamma_{ij}dx^idx^j), 
\label{bulkmetric}\end{equation}
where  $\gamma_{ij}$ is a 3D spatially flat  metric. The bulk potential corresponds to a Liouville potential; i.e.
\begin{equation} 
V(\phi)=\Lambda\exp[-(2/3) k\phi].
\label{liouville}\end{equation}
The parameters $k$ and $\xi$ in Eq.~(\ref{bulkmetric}) measure the magnitude of the 5D cosmological constant $\Lambda$ 
in the Liouville potential
\begin{equation} \Lambda= \frac12 (k^{2}-12)\xi^2. \end{equation}
On the other hand, the 5D scalar field scales logarithmically with the radial coordinate $r$ \cite{Kunze:2001ji}
\begin{equation} \phi=k\log (r). \label{phi}\end{equation}
We will consider only the case $k > 0$ (the main conclusions of the paper does not depend on the sign of $k$); i.e. the scalar field is a growing function of the coordinate $r$. The metric (\ref{bulkmetric}) has a naked singularity at $r=0$: the Ricci scalar diverges for vanishing
$r$,
\begin{equation}
R[g^{(5)}]=\frac43\,{\xi}^{2} \left( 2\,{k}^{2}-15 \right) {r}^{-(2/3)\,{k}^{2}},
\end{equation}
and the singularity is not surrounded by an event horizon. 

\subsection{The brane}

We consider a homogeneous and isotropic brane embedded in the previous 5D dilatonic solution Eq.~(\ref{bulkmetric}) 
and whose trajectory on the bulk is described by the following parametrisation 
\begin{equation}{t}={t}(\tau),\,\,\,\, r=a(\tau),\,\,\,\, x_i= \textrm{constant},\,\, i=1 \ldots 3, \end{equation}
where $\tau$ corresponds to the proper time of the brane. Then the brane metric reads
\begin{equation} 
ds^2_4\,=\,g_{\mu\nu}\,dx^{\mu}dx^{\nu}\,=\,-d\tau^2+a^2(\tau)\gamma_{ij}dx^idx^j.
\end{equation}
For simplicity we will assume a $\mathbb{Z}_2$-symmetry at the brane (which is also motivated by specific M-theory constructions \cite{HW,LOW}). For an induced gravity brane-world model \cite{Deffayet,Bouhmadi-Lopez:2004ax}, there are two physical ways of embedding the brane on the bulk when a $\mathbb{Z}_2$-symmetry across the brane is assumed. We will refer to a brane as the positive branch when the location of the brane  $r=a(\tau)$ is such that 
\begin{equation}
ds^2_5=\frac{1}{\xi^2}r^{2/3(k^2-3)}dr^2 +r^2(-d{t}^2+\gamma_{ij}dx^idx^j), \quad r>a(\tau).
\label{bulkmetric+}\end{equation}
On the other hand, we will refer to a brane as the negative branch when the location of the brane  $r=a(\tau)$ is such that 
\begin{equation}
ds^2_5=\frac{1}{\xi^2}r^{2/3(k^2-3)}dr^2 +r^2(-d{t}^2+\gamma_{ij}dx^idx^j),  \quad r<a(\tau).\label{bulkmetric-}\end{equation}

The brane tension $\lambda$ is constrained by the scalar field junction condition across the brane \cite{Chamblin:1999ya}
\begin{eqnarray} 
n^a\nabla_a\phi=\frac{\kappa_5^2}{2
}\left[\lambda'(\phi)-(\ln\Omega)'(\phi)\,T\right],\;\;\; T=T_\mu^\mu,\label{junctionphi}
\end{eqnarray}
where a prime stands for derivative respect to the scalar field
and $T_{\mu\nu}$ is defined as   
\begin{equation}
T_{\mu\nu}=-\frac{2}{\sqrt{-g}}\frac{\delta}{\delta g^{\mu\nu}}(\sqrt{-g}\,\,\Omega^4\mathcal{L}_m).
\end{equation}
We will consider that $T_{\mu\nu}$ is described effectively by a perfect fluid with energy density $\rho$ and pressure $p$. For simplicity, we will consider that the matter content of the brane is minimally coupled respect to the  conformal metric 
${\tilde g}^{(5)}_{AB}=\exp(2b\phi)\;g^{(5)}_{AB}$; 
i.e. $\Omega=\exp(b\phi)$, where $b$ is a constant.

The evolution equation for the scale factor of the brane, $a$, is  given by the Israel junction conditions \cite{Chamblin:1999ya,Maeda:2000wr} 
\begin{eqnarray} K_{\mu\nu}=-\frac{\kappa_5^2}{2}\left[S_{\mu\nu}-\frac13 S g_{\mu\nu}\right],\;\; \;\;\; S=S_\mu^\mu,\label{israel}
\end{eqnarray}
where $S_{\mu\nu}$ is the  energy-momentum tensor of the brane  
\begin{equation}
S_{\mu\nu}=-\frac{2}{\sqrt{-g}}\frac{\delta}{\delta g^{\mu\nu}}(\sqrt{-g}\,\,\mathcal{L}_4),
\end{equation}
and can be split as
\begin{equation} S_{\mu\nu}=-\lambda(\phi) g_{\mu\nu}+T_{\mu\nu} -2\alpha G_{\mu\nu}. \label{defS}\end{equation}
The last term on the rhs of the previous equation corresponds to the effect of the induced gravity term on the energy momentum tensor of the brane. This term is proportional to the Einstein tensor of the brane  $G_{\mu\nu}$.  We would like to stress that although  the induced gravity effect does not appear explicitly in the junction condition for the scalar field $\phi$ Eq.~(\ref{junctionphi})  (because there is no coupling between the scalar 
field and the induced gravity term), the brane tension will feel the induced gravity effect through the modified Friedmann equation. 

The $ij$ components of Israel junction condition Eq.~(\ref{israel}) imply:
\begin{equation} \sqrt{{\xi^2}{a^{-\frac23 k^2}}+H^2} =-\epsilon\,\frac{\kappa_5^2}{6}
\left[\lambda(\phi)+\rho-6\alpha H^2\right],
\label{Friedmann}\end{equation}
where $\epsilon=\pm 1$. From now on, $\epsilon=1,-1$ refer to the positive and negative branches, respectively. 
The $tt$ component of Eq.~(\ref{israel}) results on:
\begin{equation} \frac{\xi^2 a^{-\frac23 k^2}+(\frac{k^2}{3}+1)H^2+\dot H}{\sqrt{\xi^2{a^{-\frac23 k^2}}+H^2}}=\epsilon\,
\frac{\kappa_5^2}{2}\left\{-\frac13\lambda(\phi)+\left[\frac23\rho+p+2\alpha(H^2+2\dot H)\right]\right\},
\end{equation}
where a dot stands for derivative respect the cosmic time $\tau$.

Using the junction condition of the scalar field at the brane Eq.~(\ref{junctionphi}), we obtain
\begin{equation}
k\,\,\sqrt{{\xi^2}{a^{-\frac23 k^2}}+H^2}=\epsilon\frac{\kappa_5^2}{2}\left[\lambda'(\phi) 
-b(-\rho+3p)\right].\label{junctionphi1}
\end{equation}

The energy balance on the brane due to the scalar field comes from the Codacci equation \cite{Maeda:2000wr}, the junction conditions (\ref{junctionphi}), (\ref{israel}) and reads 
\begin{equation}
\nabla_{\nu}{S^{\nu}}_{\mu}=-(\lambda'(\phi)-b T)\nabla_{\mu}\phi.
\end{equation}
Substituting the explicit expression of the  energy-momentum of the brane Eq.~(\ref{defS}) in the previous expression, we obtain
\begin{equation}
\nabla_{\nu}{T^{\nu}}_{\mu}=b\, T\,\nabla_{\mu} \phi.
\end{equation}
Consequently
\begin{equation}
\dot\rho+3H(\rho+p)=(\rho-3p)b\,\dot a \frac{d\phi}{da}.
\end{equation}
The energy density on the brane is not conserved as is usual in dilatonic brane-world models \cite{Chamblin:1999ya,Maeda:2000wr}. Indeed,  the presence of an induced gravity term on the brane action does not modify the energy balance on the brane because the induced gravity term is not coupled to the bulk scalar field; i.e. $\alpha$ is constant.
Then, if we suppose that the perfect fluid satisfies a constant 
equation of state; i.e.  $p=w\rho$ with $w$ constant, the energy density scales with the scale factor as in the model analysed in \cite{Kunze:2001ji}:
\begin{equation}
\rho(\phi)=\rho_0\exp[-(m/k)\phi],\qquad m=3(w+1)+(3w-1)kb.\label{balance}
\end{equation}
However, the Friedmann equation of the brane is modified by the presence of  the induced gravity term as we will see shortly. 
Moreover, the differential equation satisfied by the 
brane tension $\lambda$ involves the Hubble parameter $H$ (see Eq.~(\ref{junctionphi1})). Consequently, the solutions for $\lambda(\phi)$ obtained in \cite{Kunze:2001ji} are no longer valid in our model.

Friedmann equation of the brane follows from Eq.~(\ref{Friedmann}) and reads
\begin{equation}
H^2=-\xi^2a^{-(2/3) k^2} +\frac{\kappa_5^4}{36}\left[\lambda+\rho-6\alpha H^2\right]^2.
\end{equation}
There are two solutions of the Friedmann equation
\begin{equation}
H^2=\frac{1}{6\alpha}\left\{\lambda+\rho+\frac{3}{\kappa_5^4\alpha}
\left[1+\epsilon\sqrt{1+4\kappa_5^4\alpha^2\xi^2a^{-2k^2/3}+\frac{2}{3}\kappa_5^4\alpha(\rho+\lambda)}\right]\right\},\label{Friedmann3}
\end{equation}
where $\epsilon=1$ describes the cosmological evolution of the positive branch and $\epsilon=-1$ describes the cosmological evolution of the negative branch. The positive branch has not a well defined limit 
when $\alpha$ approaches zero. This branch generalises the DGP model \cite{Dvali:2000hr}. Indeed, for $\lambda=\rho=0$ and for a constant scalar field and vanishing potential on the bulk; i.e. $k=0$ and $\xi=0$, we recover the self-accelerating solution of the DGP model \cite{Deffayet}. The negative branch  has a well defined limit 
where we recover the Friedmann equation obtained in \cite{Kunze:2001ji}. Moreover, this branch generalises the RS model \cite{RS}. In fact, for a vanishing induced gravity parameter, a constant scalar field on the bulk and a constant potential on the bulk fine tuned with the brane tension such that $36\xi^2=\kappa_5^4\lambda^2$, we recover RS model.

It remains to deduce the constraint on the brane tension.
Using equations  (\ref{Friedmann}), (\ref{junctionphi1}) and (\ref{balance}) we obtain
\begin{equation}
H^2=\frac{1}{2k\alpha}\left[(\lambda+\rho)'+\frac{k}{3}(\lambda+\rho)+\frac{3}{k}(w+1)\rho\right].
\label{Friedmann2}\end{equation}
Finally, equating the last two equations, we deduce the constraint satisfied by  the brane tension for each of the branches
\begin{equation}
(\lambda+\rho)'+\frac{3}{k}(w+1)\rho=\frac{k}{\kappa_5^4\alpha}\left[
1+\epsilon\sqrt{1+4\kappa_5^4\alpha^2\xi^2a^{-2k^2/3}+\frac23\kappa_5^4\alpha(\lambda+\rho)}\right].
\label{diflambda}\end{equation}
Again, in the limiting case where the induced gravity parameter vanishes the last equation is well defined for the negative branch. This is not true for the positive branch.

\section{Vacuum Brane solutions}

In what follows, we will restrict our analysis to the case of a vacuum brane; i.e. $p=-\rho$. Then the constraint equation of the brane tension (\ref{diflambda}) reduces to 
\begin{equation}
(\lambda+\rho)'=\frac{k}{\alpha\kappa_5^4}
\left[1+\epsilon\;\;\sqrt{1+\frac23 \alpha \kappa_5^4 (\lambda+\rho)+4\alpha^2\kappa_5^4\xi^2 a^{-(2/3) k^2}}\right].
\label{lambdaeq}\end{equation}
If there is no matter on the brane then the brane tension satisfies the same constraint Eq.~(\ref{lambdaeq}) 
with $\rho=0$. This is not surprising as we have restricted the matter content on the brane to satisfy a vacuum equation of state. 
Consequently, the energy density $\rho$ for $w=-1$ results in a rescaling of the brane tension $\lambda$. For the sake of simplicity, 
we will refer to $\lambda+\rho$ (with $w=-1$) simply as the brane tension.

Before getting the general solution of Eq.~(\ref{lambdaeq}), let us see if there are maximally symmetric branes (we restrict to Minkowski and de Sitter space-times). Combining Eqs.~(\ref{Friedmann2}) and (\ref{lambdaeq}), it can be shown that there are maximally symmetric branes as long as the brane tension scales with the scale factor as
\begin{equation}
\lambda= 6\alpha H^2 -\epsilon\frac{6}{\kappa_5^2}\sqrt{H^2+\xi^2a^{-\frac23 k^2}}\label{lambdadesitter}
\end{equation}
and satisfy the differential equations (\ref{Friedmann2}) and (\ref{lambdaeq}). This is only possible for a flat brane; i.e. $H=0$. Indeed in this case the brane tension is fine-tuned with the cosmological constant. This kind of solutions was already noticed in \cite{Kachru:2000hf}. A flat brane in the negative branch can be embedded in the bulk given by Eq.~(\ref{bulkmetric}) for any value of the scalar field, $\phi$, at the brane. However, a flat brane in the positive branch can only be embedded on the bulk, if $\phi\leq3/(2k)\log(d)$ (see Eqs.(\ref{Friedmann3}) and (\ref{lambdadesitter})). 

There are no brane solution with de Sitter geometry for an evolving bulk scalar field. However, Eq.~(\ref{lambdadesitter}) satisfies Eqs.~(\ref{Friedmann2}) and (\ref{lambdaeq})  for very small scale factors on both branches; i.e. the positive and negative branches can be asymptotically de Sitter in the past. On the other hand, it can also be checked that the negative branch is asymptotically Minkowski in the future; i.e. Eq.~(\ref{lambdadesitter}) satisfies Eqs.~(\ref{Friedmann2}) and (\ref{lambdaeq})  for very large scale factors.

In order to solve the differential  equation (\ref{lambdaeq}) 
it is helpful to rewrite it as:
\begin{equation}
\frac{dy}{dx}=1+\epsilon\;\;\sqrt{1+y+de^{-x}},
\label{dify}\end{equation}
in terms of the following dimensionless quantities 
\begin{equation}
y=\frac23 \kappa_5^4 \alpha \lambda,\quad x=\frac23 k\phi=\frac23k^2\ln(a),\quad d= 4\alpha^2\kappa_5^4\xi^2.
\label{def}\end{equation}
The solution of the  previous equation can be written as a 2D surface involving the dimensionless brane tension $y$ and  the rescaled dimensionless scalar field $x$
\begin{equation}
2\sqrt{z^2-de^{-x}}+\ln\left[\frac{z-\sqrt{z^2-de^{-x}}}{z+\sqrt{z^2-de^{-x}}}\right]=2c,\quad z=1+\epsilon\sqrt{1+y+de^{-x}},
\label{surface}\end{equation}
where $c$ is an integration constant. We will show latter on that the initial value of the Hubble parameter is fixed by $c$. The Minkowski solution we discussed earlier corresponds to $z^2=d\exp(-x)$ or equivalently $y=2\epsilon\sqrt{d}\exp(-x/2)$. 

Although, we are not able to get an explicit relation between the brane tension and the scale factor (or between $y$ and $x$), we can rewrite the constraint (\ref{surface}) in a parametric form as\footnote{The parametric solution given by Eqs.~(\ref{+solution1})-(\ref{+solution}) does not include the Minkowski solution, we discuss earlier.} 
\begin{eqnarray}
z(\eta)&=&\frac{1}{2\eta}\left[2c-\ln\left(\frac{1-\eta}{1+\eta}\right)\right],\label{+solution1}\\
de^{-x}(\eta) &=&(1-\eta^2)z^2,\label{xeta}\label{+solution2}\\
y(\eta)&=&z(-2+\eta^2 z).
\label{+solution}\end{eqnarray}

Finally, substituting Eqs.~(\ref{+solution1})-(\ref{+solution}) into  Friedmann equation (\ref{Friedmann3}), the Hubble parameter reads:
\begin{equation}
 H= \frac{1}{\sqrt{2}\alpha\kappa_5^2}\, h=\frac{1}{2\alpha\kappa_5^2}\,\eta z,\label{defh}
\end{equation}
where, for latter convenience, we have introduced the dimensionless Hubble parameter, $h$, which in term of $\eta$ reads
\begin{equation}
h=\frac{\sqrt{2}}{4}\left[2c-\ln\left(\frac{1-\eta}{1+\eta}\right)\right].
\label{defh1}\end{equation}

The range of variation of the parameter $\eta$ and the constant $c$ depends crucially on the specific branch. This will result in very different physical properties of the two branches.

\subsection{Positive branch $\epsilon=1$}

In the positive branch and for a vacuum brane, the brane tension is a growing function of the scalar field or equivalently of the scale factor of the brane (see Eqs.~(\ref{phi}) and (\ref{junctionphi1})). We would like to stress that in a standard 4D FLRW expanding universe filled with a  phantom energy component, the energy density behaves in this way too: the energy density is a growing function of the scale factor \cite{Caldwell:2003vq,Bouhmadi-Lopez:2004me}.

The dimensionless variable $z$ which essentially measures the growth of the brane tension with respect to the scalar field is greater than 1 (see Eq.~(\ref{surface})). This implies $\eta\in[0,1)$ and the integration constant c is such that $c\geq 0$. In addition, there is a one to one correspondence between the scale factor and the parameter $\eta$. Indeed, the scale factor is an increasing function\footnote{The function $de^{-x}=da^{-2/3 k^2}$ is a decreasing function of $\eta$.\label{footnotea}} of $\eta$.

A detailed analysis of Eqs.~(\ref{+solution1})-(\ref{+solution}) for $c>0$ shows that at high energy ($a \rightarrow 0$ or $\eta\rightarrow 0$), the brane tension reaches infinite negative values. On the other hand, for very large values of the scale factor ($\eta\rightarrow 1$), the brane tension acquires infinite positive values. The larger is the value of the integration constant $c$, the sooner (for smaller scale factor) the brane tension becomes positive. 

The behaviour of the solution given by Eqs.~(\ref{+solution1})-(\ref{+solution}) for $c=0$ 
is different from the behaviour of the solution with $c>0$. The scale factor is bounded from below (for $\eta\rightarrow 0$,  $d\exp(-x)\rightarrow 1$) when $c=0$.  The brane tension is also bounded from below (for $\eta\rightarrow 0$, $y\rightarrow -2$). On the other hand, the brane tension acquires very large positive values when the scale factor is very large (or $\eta\rightarrow1$). Moreover, the solution given by Eqs.~(\ref{+solution1})-(\ref{+solution}) for $c=0$ can be extended to $-1<\eta<0$. In fact, the  solution is bouncing around $\eta=0$ where the scale factor acquires its minimum value. For simplicity, we will restrict to $0\leq\eta<1$ when $c=0$.

The dimensionless Hubble parameter $h$ is an increasing function of $\eta$ and consequently an increasing function of the scale factor $a$ (see footnote \ref{footnotea}).
At high energy (small values of the scale factor or  $\eta$ approaching $0$), the dimensionless parameter $h$ reaches  a constant positive value proportional to the integration constant $c$. 
On the other hand, at very large values of the scale factor (or $\eta$ approaching $1$), the dimensionless Hubble parameter diverges.
The divergence of the Hubble parameter for very large values of the scale factor  might point out the existence of a big rip singularity in the future evolution of the brane; i.e. the scale factor and Hubble parameter blow up in a finite cosmic time in the future evolution of the brane. However, as we will next show the divergence of $H$  and  $a$ (and also of $\lambda$) occur in an infinite cosmic time in the future evolution of the positive branch. 

We introduce a dimensionless cosmic time for the brane defined  as 
\begin{equation}
\tilde\tau=\frac{1}{2\alpha\kappa_5^2}\,\tau.
\end{equation}
Then using Eqs.~(\ref{def}),~(\ref{xeta}) and (\ref{defh}), we obtain
\begin{eqnarray}
\tilde\tau&=&-\frac{3}{2k^2}\int_{\eta_c}^\eta \frac{g(\eta)}{(1-\eta)\ln(1-\eta)}d\eta, 
\end{eqnarray}
where
\begin{eqnarray}   
g(\eta)&=& \frac{1}{\eta z}(1-\eta)\ln(1-\eta)
\frac{d}{d\eta}\ln\left[(1-\eta^2)z^2\right],
\end{eqnarray}
and $\eta_c$ is a constant. The function $g(\eta)$ is continuous on $[0,1]$, 
approaching 2 for large value of the scale factor, i.e. $\eta=1$, and approaching $2/c$
for very small scale factor \footnote{For $c=0$, $g(\eta)$ vanishes when $\eta$ approaches zero.}; i.e. $\eta=0$. 
Consequently, $g(\eta)$ has a minimum on $[0,1]$ which we denote $\textrm{min}(g)$. On the other hand, it 
can  be proven 
that $g(\eta)$ is positive and different from zero on  $[0,1]$ for any $c>0$. 
Then $\textrm{min}(g)>0$. Consequently, the dimensionless 
cosmic time of the brane is bounded from below as follows
\begin{equation}
\tilde\tau\,>\,-\frac{3}{2k^2}\textrm{min}(g)\int_{\eta_c}^\eta\left[(1-\eta)\ln(1-\eta)\right]^{-1}d\eta.
\end{equation}
Integrating  the rhs of the previous inequality, it can be easily proven that the cosmic time on the brane diverges 
for $\eta$ approaching~1 where the scale factor reaches infinite values. 
This proof is only valid for $c>0$. For $c=0$ a straightforwards extension of the previous proof implies the same result; 
i.e. $\tilde\tau$ 
diverges when $\eta$ approaches~1 where the scale factor and Hubble parameter blow up. Another way of showing that there is no big rip singularity on the future evolution of the brane is by  noticing that the asymptotic behaviour of the dimensionless tension of the brane  at large value of the scale factor is $y\sim x^2/4$. Therefore, 
\begin{equation}
H\sim\frac{k^2}{\kappa_5^2 \alpha}\ln (a).
\end{equation}
Consequently, the Hubble rate does not grow as fast as in phantom energy models with a constant equation of state where a big rip singularity takes place on the future evolution of a homogeneous and isotropic universe \cite{Caldwell:2003vq,Bouhmadi-Lopez:2004me}. 

In summary, we have proven that in the positive branch filled only with vacuum energy there is a singularity in the future evolution 
of the brane. The singularity is  such that for large value of the cosmic time, the scale factor and the Hubble parameter diverge. 
This kind of singularity 
can be interpreted  as a ``big rip''  singularity pushed towards an infinite cosmic time of the brane. 
We would like to stress that this 
singularity appears only on the brane and not in the bulk. Indeed, large values of the scale factor, $a$, correspond to large values of the 
extra coordinate $r$ (where the bulk is asymptotically flat) and the only bulk  singularity is located at $r=0$. It can also be proven that the
 time derivative of the Hubble parameter,
\begin{equation}
2\sqrt{2}\alpha^2\kappa_5^4\dot H =\frac{dh}{d\tilde\tau}=\frac{dh}{d\eta}\left(\frac{d\tilde\tau}{d\eta}\right)^{-1}=
-\frac{\sqrt{2}k^2}{12}\,\eta \left[ 2\,c-\ln  \left( {\frac {1-\eta}{1+\eta}} \right)  \right]
^{2} \left[ 2\,\eta-2\,c+\ln  \left( {\frac {1-\eta}{1+\eta}} \right) 
 \right] ^{-1},
\label{hdot}\end{equation}
acquires infinite positive  values for very large scale factors or cosmic time. In fact, the whole evolution of the positive branch is super-inflationary, i.e. $\dot H >0$. The curvature  singularity we have found has some similarities with the one in 4D phantom  energy models with 
constant equation of state 
\cite{Caldwell:2003vq,Bouhmadi-Lopez:2004me}: $H$ and $\dot H$ diverges for large values of the scale factor. Although, in
these models \cite{Caldwell:2003vq,Bouhmadi-Lopez:2004me} the scale factor, Hubble parameter and cosmic time derivative of the Hubble parameter diverge on a finite future cosmic time. 

Finally, we point out that there is no big bang singularity for the solutions we have analysed. Despite the fact that the scale factor vanishes at earlier time for $c>0$, the Hubble parameter is finite and its time derivative vanishes (see Eq.~(\ref{hdot})). Indeed, for $c>0$ the brane geometry is asymptotically de Sitter in the past. For $c=0$, the initial scale factor is non vanishing, the Hubble parameter vanishes and its time derivative is finite. As we mentioned before, the scale factor is bouncing  around $\eta=0$ when $c=0$.

\subsection{Negative branch $\epsilon=-1$}

In the negative branch and for a vacuum brane, the brane tension is positive and a decreasing function of the scalar field or equivalently of the scale factor of the brane (see Eqs.~(\ref{phi}), (\ref{Friedmann}) and (\ref{junctionphi1})). Consequently, the dimensionless variable $z$ defined in Eq.~(\ref{surface}) is negative. This implies $\eta\in(-\tanh(c),0]$ and the integration constant satisfies $c>0$.  
It can be shown that there is a one to one correspondence between the scale factor and the parameter $\eta$. Indeed, the scale factor is a decreasing function\footnote{The function $de^{-x}=da^{-2/3 k^2}$ is a increasing function of $\eta$ and consequently the scale factor is a decreasing function of $\eta$ (see Eq.~(\ref{def})).\label{footnotec}} of $\eta$. An analysis  of Eqs.~(\ref{+solution1})-(\ref{+solution}) shows that at high energy ($a \rightarrow 0$ or $\eta\rightarrow 0$), the brane tension reaches infinite positive values. On the other hand, for very large value of the scale factor; i.e. $\eta\rightarrow -\tanh(c)$, the brane tension vanishes.

The dimensionless Hubble parameter $h$  given in  Eq.~(\ref{defh1}) is an increasing function of $\eta$ and consequently a decreasing function of the scale factor.
At high energy, the dimensionless Hubble parameter $h$ reaches  a constant positive value proportional to the integration constant $c$. 
On the other hand, at very large values of the scale factor, the dimensionless Hubble parameter vanishes.

\begin{figure}[t]
\includegraphics[width=0.8\textwidth]{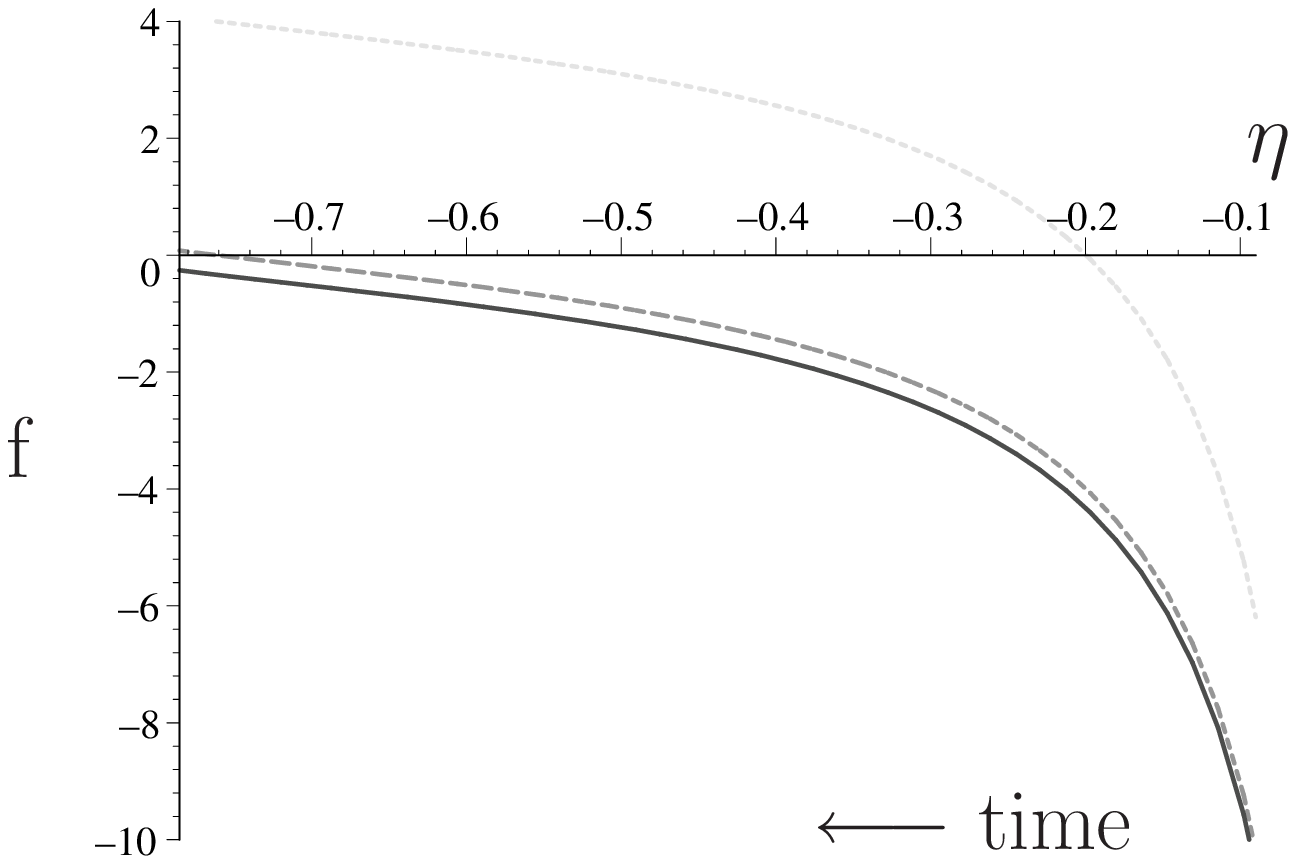}
\caption{This figure shows the behaviour of the function $f$ (defined in Eq.~(\ref{feta})) as a function of $\eta$. The negative branch undergoes an inflationary expansion whenever $f$ is negative. For $k^2\leq 3$, the negative branch is eternally inflating. On the other hand, for $k^2>3$ the brane is initially inflating until $f(\eta)$ vanishes. The solid (darker grey), dotted and dashed-dotted (lighter grey) lines correspond to $f(\eta)$ for $k^2=2,3,15$ respectively and $c=1$.}
\label{inflation}
\end{figure}

\begin{figure}[b]
\includegraphics[width=0.8\textwidth]{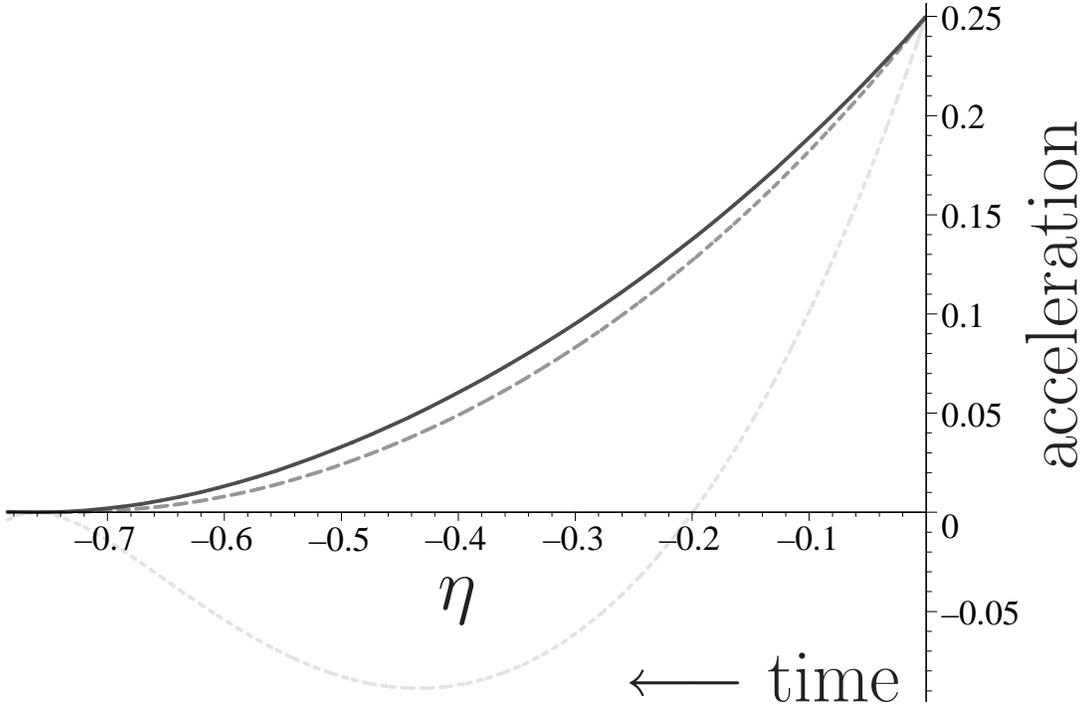}
\caption{This figure shows the behaviour of the dimensionless  acceleration parameter given by $\alpha^2\kappa_5^4\ddot{a} /a$ (introduced in Eq.~(\ref{adotdot})) as a function of the parameter $\eta$. The solid (darker grey), dotted and dashed-dotted (lighter grey) lines correspond to the acceleration parameter for $k^2=2,3,15$ respectively and $c=1$. For $k^2= 2,3$ the negative branch is eternally inflating. On the other hand, for $k^2= 15$ the brane undergoes an initial transient inflationary epoch.}
\label{inflation2}
\end{figure}

Unlike the positive branch, the negative branch is never super-inflating, although the negative branch always undergoes an inflationary period, as we will next show.  The second temporal derivative of the scale factor satisfies

\begin{equation}
\alpha^2\kappa_5^4\frac{\ddot a}{a}=-\frac{1}{2}\,\frac{\eta z}{1-\eta^2}\left[\frac{d}{d\eta}\ln\left(d e^{-x}\right)\right]^{-1}f(\eta),
\label{adotdot}\end{equation}
where 
\begin{equation}
f(\eta)=z(\eta)-1+\frac{k^2}{3}.
\label{feta}
\end{equation}
The brane is inflating when $f$ is negative\footnote{The term multiplying $f$ on the rhs of Eq. (\ref{adotdot}) is negative because $de^{-x}$ is positive and a decreasing function of $\eta$, $-1<-\tanh (c)<\eta\leq 0$ and the brane tension is positive (z is negative). Consequently, the brane is inflating whenever $f$ is negative.}. The brane behaves in two different ways depending on the values acquired by $k^2$. For $k^2\leq 3$, the brane is eternally inflating as $f$ is always negative (see Figs.~\ref{inflation},\ref{inflation2}). A similar behaviour was found in \cite{Kunze:2001ji} in a dilatonic brane-world model without an induced gravity term. On the other hand, the brane undergoes an initial stage of inflation when $k^2>3$ until $f$ vanishes. Later on the brane starts decelerating (see Figs.~\ref{inflation},\ref{inflation2}). This second behaviour was not found in \cite{Kunze:2001ji} for a vacuum brane. Then, the inclusion of an induced gravity term on a dilatonic brane-world model with an exponential potential on the bulk allows inflation to take place in the negative branch, even if it does not for a vanishing induced gravity parameter (for $k^2>3$). This behaviour has some similarity with steep inflation \cite{Copeland:2000hn}, where high energy corrections to the Friedmann equation in RS scenario \cite{RS} permit an inflationary evolution of the brane with potentials too steep to sustain it in the standard 4D case, although the inflationary scenario introduced by Copeland et al in  \cite{Copeland:2000hn} is supported by an inflaton confined in the brane while in our model inflation on the brane is induced by a dilaton field on the bulk.

Finally, we stress that there is no big bang singularity 
for the vacuum solution on the negative branch.
Although, the scale factor vanishes at early times, the Hubble parameter is finite and its time derivative vanishes (see Eq.~(\ref{hdot})). Like the positive branch, the negative branch is asymptotically de Sitter in the past. The vacuum negative branch  also does not  face a curvature singularity in its future. Indeed, for large values of the scale factor the Hubble parameter and its cosmic derivative vanish and the brane is asymptotically Minkowski.

\section{Conclusions}

In this paper we study the behaviour of a dilatonic brane-world model with an induced gravity term on the brane with a constant induced gravity parameter $\alpha$. We assume a  $\mathbb{Z}_2$-symmetry across the brane. The dilatonic potential is an exponential function of the bulk scalar field and the matter content of the brane is coupled to the dilaton field. 
We deduce the modified Friedmann equation for the positive and negative branch (which specifies the way the brane is embedded in the bulk), the junction condition for the scalar field across the brane and the energy balance on the brane. We obtain the vacuum solutions Eqs.~(\ref{+solution1})-(\ref{+solution}); i.e. the matter content of the brane is specified by the brane tension, for a FLRW brane.

In the vacuum positive branch, the brane tension is a growing function of the scale factor and, consequently, mimics the behaviour of a phantom energy component on the brane. This phantom-like behaviour is obtained without including a phantom fluid on the brane. In fact, the brane tension does not violate the null energy condition. The expansion of the brane is super-inflationary; i.e. the Hubble parameter is a growing function of the cosmic time. At high energy (small scale factors), the brane is asymptotically de Sitter  for $c>0$, where the parameter $c$ is proportional to the initial Hubble parameter. There is a another solution which is a  bouncing solution ($c=0$) around a minimum non-vanishing scale factor. The brane faces a curvature singularity in its future evolution, where the Hubble parameter, brane tension and scale factor diverge. The singularity happens in an infinite cosmic time. Therefore, the singularity can be interpreted as a ``big rip'' singularity pushed towards an infinite future cosmic time.

In the vacuum negative branch, the brane tension is a decreasing function of the scale factor. Unlike the positive branch, the branch is not super-inflating. However, it always undergoes an inflationary expansion (see Eq.~(\ref{adotdot})). The inflationary expansion can be eternal ($k^2\leq 3$) or transient ($k^2> 3$). For large values of the scale factor, the negative branch is asymptotically Minkowski.

It remains to be seen how the  results obtained in this paper are modified  when the brane has  matter contents in addition to  the brane tension. On the other hand, we have chosen a specific dilatonic bulk and consequently we do not know how general are the brane properties we have found. We leave these interesting questions for future work.

\acknowledgments

We thank D. Wands for very enlightening discussions. The author is also grateful to A. Ferrera, P. F. Gonz\'{a}lez-D\'{\i}az, K. Koyama and R. Maartens for comments.  The author thanks R.~Lazkoz for encouragements.
MBL acknowledges the support of FCT (Portugal) through the fellowship SFRH/BPD/26542/2006 and the research grant FEDER-POCI/P/FIS/57547/2004. The initial work of MBL was funded by MECD (Spain) and also partly supported by DGICYT under Research Project BMF2002 03758. This paper is dedicated to all the people who helped me to recover the mobility of my arm, in particular, the traumatology team of San Juan de la Cruz Hospital, Cati, Conchi and Juana.

\newpage

\end{document}